\documentclass[sigconf, screen, nonacm]{acmart}

\AtBeginDocument{%
  \providecommand\BibTeX{{%
    \normalfont B\kern-0.5em{\scshape i\kern-0.25em b}\kern-0.8em\TeX}}}

\acmConference[]{}{}{}

\usepackage[shortcuts]{extdash}

\usepackage{tikz}
\usepackage{enumitem}
\setlist{itemsep=2pt plus 2pt minus 1pt,topsep=3pt plus 2pt minus 1pt} 
\begin{document}

\newcommand{\code}[1]{\texttt{#1}}
\newcommand{\coolname}{dMVX}
\newcommand{\oldtoolname}{DMON}
\newcommand{\discrossprocessmonitor}{DCP\=/MON}
\newcommand{\disinprocessmonitor}{DIP\=/MON}
\newcommand{\connector}{connector}
\newcommand{\broker}{SA}
\newcommand{\cb}{CB}

\newcommand*\circledb[1]{\tikz[baseline=(char.base)]{
    \node[shape=circle,draw,inner sep=1pt,fill=black] (char)
         {\textcolor{white}{\footnotesize #1}};}}
\newcommand*\circledbw[1]{\tikz[baseline=(char.base)]{
    \node[shape=circle,draw,inner sep=1pt,fill=white] (char)
         {\textcolor{black}{\scriptsize #1}};}}
\newcommand*\circledbg[1]{\tikz[baseline=(char.base)]{
    \node[shape=circle,draw,inner sep=1pt,fill=gray] (char)
         {\textcolor{black}{\scriptsize #1}};}}
\newcommand*\circledblg[1]{\tikz[baseline=(char.base)]{
    \node[shape=circle,draw,inner sep=1pt,fill=gray90] (char)
         {\textcolor{black}{\scriptsize #1}};}}

\newcommand*\dcircledw[1]{\tikz[baseline=(char.base)]{
		\node[shape=circle,draw,inner sep=.75pt, dashed, dash pattern = on 1.5pt off .75pt, line width=.7pt] (char) {\scriptsize #1};}}

\usetikzlibrary{patterns}

\newcounter{dscnt}
\newcounter{avcnt}
\newcounter{svcnt}
\newcounter{adcnt}

\newcommand{\ds}[1]{\refstepcounter{dscnt}
  \textcolor{red}{\textbf{Dokyung [\thedscnt]:} #1}}
\newcommand{\av}[1]{\refstepcounter{avcnt}
  \textcolor{purple}{\textbf{Alex [\theavcnt]:} #1}}
\newcommand{\sv}[1]{\refstepcounter{svcnt}
  \textcolor{green}{\textbf{Stijn [\thesvcnt]:} #1}}
\newcommand{\ad}[1]{\refstepcounter{adcnt}
  \textcolor{red}{\textbf{Adrian [\theadcnt]:} #1}}

\title[]{\coolname{}: Secure and Efficient Multi-Variant Execution in a Distributed Setting}


\author{Alexios Voulimeneas}
\affiliation{
  \institution{imec-DistriNet, KU Leuven}
}
\email{alex.voulimeneas@kuleuven.be}

\author{Dokyung Song}
\affiliation{
  \institution{UC Irvine}
}
\email{dokyungs@uci.edu}

\author{Per Larsen}
\affiliation{
  \institution{UC Irvine}
}
\email{perl@uci.edu}

\author{Michael Franz}
\affiliation{
  \institution{UC Irvine}
}
\email{franz@uci.edu}

\author{Stijn Volckaert}
\affiliation{
  \institution{imec-DistriNet, KU Leuven}
}
\email{stijn.volckaert@kuleuven.be}

\renewcommand{\shortauthors}{}

\begin{abstract}
Multi-variant execution (MVX) systems amplify the effectiveness of software diversity techniques.
The key idea is to run multiple diversified program variants in lockstep while providing them with the same input and monitoring their run-time behavior for divergences.
Thus, adversaries have to compromise all program variants simultaneously to mount an attack successfully.
%
%

Recent work proposed distributed, heterogeneous MVX systems that leverage different ABIs and ISAs to increase the diversity between program variants further.
However, existing distributed MVX system designs suffer from high performance overhead due to time-consuming network transactions for the MVX system's operations.

This paper presents \coolname{}, a novel hybrid distributed MVX design, which incorporates new techniques that significantly reduce the overhead of MVX systems in a distributed setting. Our key insight is that we can intelligently reduce the MVX operations that use \emph{expensive} network transfers. First, we can limit the monitoring of system calls that are not security-critical. Second, we observe that, in many circumstances, we can also safely cache or avoid replication operations needed for I/O related system calls.
Our evaluation shows that \coolname{} reduces the performance degradation from over 50\% to 3.1\% for realistic server benchmarks. 

\end{abstract}

\maketitle

\section{Introduction}

Memory-unsafe languages such as C and C++ are still popular choices for systems programming. Memory errors, inherent in these languages, afford bad actors opportunities for exploitation. Attackers typically take advantage of memory errors to launch code-reuse or data-only attacks~\cite{snow.etal+13, Shacham10, Shacham2007, dataonly, hu2015automatic, hu2016data}, in order to compromise their targets and/or leak secrets. While memory safety techniques could potentially protect against these vulnerabilities, they come at the cost of performance and compatibility issues~\cite{nagarakatte2009softbound, nagarakatte2010cets, song2019sanitizing}.


Since 2006, many multi-variant execution (MVX) systems have been proposed for security and reliability purposes~\cite{berger2006diehard,dieharder,bruschi2007diversified,cox2006n,salamat2009orchestra,volckaert2012ghumvee,koning2016secure,volckaert2016cloning,volckaert2016secure,xu2017bunshin,lu2018stopping,hosek2013safe, hosek2015varan, maurer2012tachyon, kim2015dual, kwon2016ldx, osterlund2019kmvx}.
The main idea is to execute diversified variants of the same program in parallel, while providing them with the same inputs and monitoring their behavior.
When the variants' behavior diverges, security-oriented MVX systems terminate them to contain the damage of a possible attack.
Multi-core CPUs' availability makes MVX systems increasingly attractive, since variants can run at native speed (assuming the host has enough idle CPU cores and memory bandwidth).

The majority of existing MVX systems execute variants on the same hardware and platform. However, attackers have managed to bypass these systems with modern code-reuse and data-only attacks that exploit the limited variant entropy space (and, therefore, diversity) on homogeneous platforms~\cite{bdb9f72346e94b59a6cd56b83f7c59b0, hu2016data}. To increase the diversity, researchers proposed distributed MVX systems that execute variants on different heterogeneous physical machines~\cite{DBLP:conf/dimva/VoulimeneasSPNL20, xiaoguang:raid20}, leveraging heterogeneous Instruction Set Architectures (ISAs) and Application Binary Interfaces (ABIs). However, existing distributed MVX systems suffer from large performance overhead.

We propose a new, distributed MVX design, \coolname{}, which incorporates new techniques that substantially reduce \coolname{}'s overhead relative to existing distributed MVX systems. The major bottleneck for these distributed systems is the \emph{expensive} (i.e., time-consuming) network communication needed for basic MVX operations. Our key insight is that many of these \emph{expensive} MVX operations can either be executed asynchronously or omitted altogether. 

\coolname{} uses an existing, security-oriented distributed MVX system (\oldtoolname{}~\cite{DBLP:conf/dimva/VoulimeneasSPNL20}) to handle security-critical (or sensitive) system calls. We also added a security-hardened distributed in-process monitor (\disinprocessmonitor{}) that enables efficient monitoring and replication of non-sensitive system calls. This split-monitor (or hybrid) design provides security guarantees that are comparable to those of existing distributed MVX systems, while outperforming them in terms of run-time overhead. ReMon proposed a similar approach~\cite{volckaert2016secure}, but it did not have to deal with the unique challenges that arise in a distributed setting (see \autoref{sec:design} for details).



Overall, our paper contributes the following:

\begin{itemize}

\item \textbf{Novel MVX Design.}
%
\coolname{} adopts a hybrid monitor design. In contrast to previous work~\cite{volckaert2016secure}, \coolname{} executes variants in a distributed setting. We describe the unique challenges that arise from a distributed setting and describe our solutions. To our knowledge, \coolname{} is the first security-focused distributed MVX design that can run and monitor variants on different physical machines at near-native speed.


\item \textbf{Efficient Monitoring and Replication.} We demonstrate techniques that reduce the performance overhead of monitoring and replication in a distributed MVX setting.

\item \textbf{Performance Evaluation.} We build a prototype and evaluate it on microbenchmarks and \texttt{lighttpd}. Our results show that \coolname{} runs realistic server benchmarks at almost native speed (3.1\% run-time overhead), significantly reducing the overhead of existing work that reports performance degradation of at least 50\%.

\end{itemize}

\section{Background}


Every MVX system contains at least one monitor component, which feeds the variants with the same inputs and observes their run-time behavior. The monitor's design characteristics are crucial for the MVX system's security and performance.

First, we can categorize monitor designs into user-space and in-kernel designs~\cite{cox2006n}. In general, an in-kernel design provides the best performance. However, it requires significant modifications to the kernel or kernel modules, which are both difficult to maintain over time and increase the size of the Trusted Computing Base (TCB). Furthermore, attacks targeting the MVX monitor itself could potentially compromise the entire system.

Second, user-space monitors can run as a standalone process (cross-process)~\cite{cavallaro_phd_thesis_2007, salamat2009orchestra, volckaert2016cloning}, inside the running variants (in-process)~\cite{hosek2015varan}, or both (hybrid)~\cite{volckaert2016secure}. Cross-process designs provide the strongest security guarantees since the operating system's process isolation protects the monitor. Several previous works used the \texttt{ptrace} API, a debugging infrastructure provided by the Linux kernel, to build their cross-process monitors~\cite{volckaert2012ghumvee, cavallaro_phd_thesis_2007, maurer2012tachyon, hosek2013safe, salamat2009orchestra, salamat2011}. However, this comes at the cost of expensive context switching~\cite{volckaert2016secure}. On the other hand, in-process designs are efficient, but provide weaker security guarantees, unless hardware~\cite{10.1145/3380786.3391398}, and/or software protection (e.g., CFI~\cite{abadi2005control} and SFI~\cite{wahbe1994efficient}) is added. Finally, Volckaert et al. proposed a hybrid design that consists of both an in-process and a cross-process monitor component, which unify the security and performance properties of previous approaches~\cite{volckaert2016secure}.

\begin{figure}[t]
	\centering
	\includegraphics[scale=0.40]{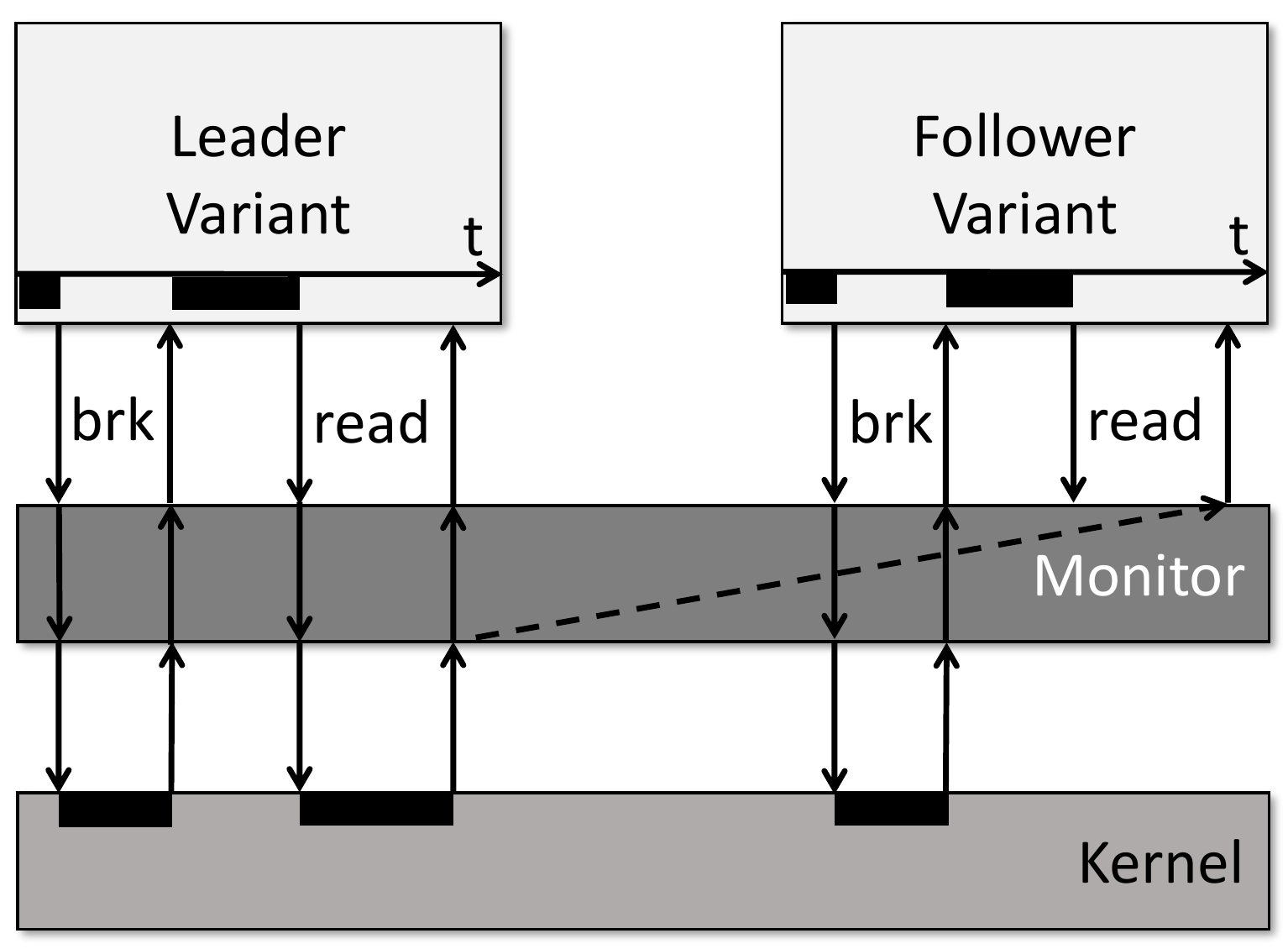}
	\caption{Leader/follower I/O replication model.}
	\label{fig:leaderfollower}
\end{figure}

\subsection{MVX Execution Model}

The majority of MVX systems perform monitoring and I/O replication operations at the system call level.
Previous work has shown that monitoring the system call interface is sufficient to provide the necessary security guarantees since all I/O operations can be monitored and replicated, and all potentially dangerous operations can be stopped at this interface~\cite{cox2006n, hosek2015varan, salamat2009orchestra}. Monitoring and replication operations should be transparent to the end-user and to the executing variants. To this end, most MVX systems adopt a leader/follower replication model, as depicted in \autoref{fig:leaderfollower}.
One variant is designated as the leader and performs all I/O related system calls, e.g., \texttt{read}, while the monitor copies the results from the leader variant to the rest (followers). On the other hand, all variants execute system calls that are not related to I/O, e.g., \texttt{brk}.


\subsection{MVX Security}

An MVX system's precise security guarantees depend on which diversity transformations it applies to each variant.
Typically, the system administrator chooses transformations that ---with high likelihood--- make the same exploit payload trigger asymmetric responses across variants.
For example, to stop traditional code-reuse attacks, the MVX system could generate one variant that uses only the upper half of the virtual address space, while the other variants use only the lower half.
With such a setup, any exploit payload that compromises one variant will simultaneously crash at least one other variant (since it will attempt to execute code on unmapped memory pages).
The MVX system detects this divergent behavior and takes appropriate action.
Similar techniques can be applied to thwart data-oriented attacks that leak or corrupt information.

However, researchers showed that MVX systems running on single-ISA machines are still vulnerable to modern code-reuse and data-only attacks~\cite{bdb9f72346e94b59a6cd56b83f7c59b0, hu2016data}, even if they apply all state-of-the-art diversity transformations. For example, Goktas et al. demonstrated practical code-reuse attacks that use relative locations of code gadgets to build exploits~\cite{bdb9f72346e94b59a6cd56b83f7c59b0}. These \emph{Position-Independent Code-Reuse Attacks} can bypass several code diversity schemes, even when augmented by MVX systems. The reason is that attackers could locate gadgets, chain them together, and build end-to-end exploits even in diversified programs. 

\subsection{Distributed MVX Systems}

Distributed MVX systems can defend against the aforementioned attacks by leveraging the additional entropy and diversity available on heterogeneous-ISA platforms~\cite{xiaoguang:raid20, DBLP:conf/dimva/VoulimeneasSPNL20}.
This additional diversity significantly increases the likelihood of exploit payloads triggering asymmetric responses in the variants, all but eliminating the available data-only and code-reuse gadgets attackers can use to construct exploits.
Two distributed MVX systems have been proposed so far.
HeterSec is built on top of a modified compiler and operating system and operates entirely in kernel space~\cite{xiaoguang:raid20}, while \oldtoolname{} operates in user space as a cross-process MVX system~\cite{DBLP:conf/dimva/VoulimeneasSPNL20}.
Both systems run variants on separate single-ISA machines connected using a fast, low-latency communication channel.
However, both suffer from high performance overhead, making them impractical for real-world workloads on system call dominant applications such as web servers. On \texttt{lighttpd}, for example, HeterSec and \oldtoolname{} report an overhead of \char`\~50\% and \char`\~443\%, respectively~\cite{xiaoguang:raid20, DBLP:conf/dimva/VoulimeneasSPNL20}.

\section{Threat Model}
\label{sec:threat_model}

We assume an attacker that can only interact with the public remote communication interface of the leader variant and has no access to the private connections between the leader and the follower(s).  Consequently, attacks targeting these private connections are out of scope of this paper.
The attacker uses said remote communication interface to send exploits, for example targeting arbitrary memory read/write vulnerabilities or control flow diversion. 


\section{Design and Implementation}
\label{sec:design}

\begin{figure}[t]
	\centering
	\includegraphics[scale=0.45]{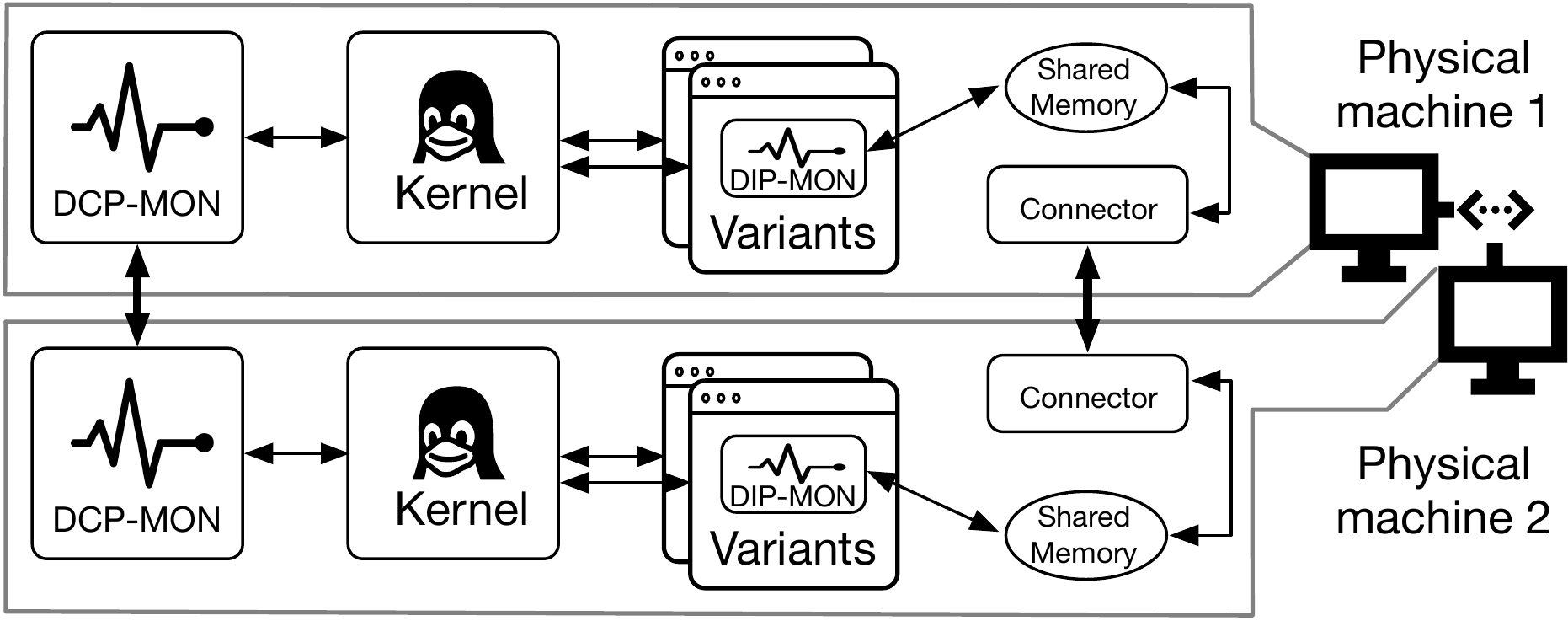}
	\caption{\coolname{'s} design.}
	\label{fig:design}
\end{figure}

We present \coolname{}, a distributed MVX system that is more efficient than prior work, while providing similar security guarantees. \autoref{fig:design} shows the design of \coolname{}. We based our design on an existing distributed MVX system~\cite{DBLP:conf/dimva/VoulimeneasSPNL20}, but added an \emph{in-process} monitoring component to more efficiently execute system calls that are not security critical. Previous work proposed a similar \emph{hybrid} MVX system design that uses both cross-process and in-process monitors~\cite{volckaert2016secure}. However, adopting this hybrid design in a distributed setting poses new challenges.

In a non-distributed MVX system, the in-process monitors can communicate directly using shared memory.
However, in a distributed MVX system, the in-process monitors have to communicate over a network connection.
Doing so forces them to issue system calls.
Unfortunately, these system calls could be inadvertently intercepted and monitored by our cross-process monitors, causing observable divergences due to the asymmetry of the sending and receiving part of the inter-monitor communication.
Consequently, one challenge to building our system is to avoid inadvertently intercepting these system calls issued by our in-process monitors, while monitoring the ones that are directly issued by the variants.

A naive approach is to add two \emph{special} system calls to enable/disable system call monitoring. The idea is to disable system call monitoring before we use the network channel for inter-monitor communication, avoiding undesired monitoring that leads to false alarms, and re-enable it again when the data exchange is finished. Note that the two \emph{special} system calls are not monitored and should never be called by the variant itself. 
However, this design is unsuitable for efficiency, since it adds four mode switches for each disable/re-enable pair. 

\subsection{System Overview}


\coolname{} supervises program variants that run in parallel on different physical machines. Its main goals
are to (i) monitor all of the security-sensitive system calls issued by the variants, (ii) execute \emph{sensitive} system calls in lockstep and (iii) relax monitoring and lockstepping for system calls that are not security critical, minimizing the overhead of our MVX system for these system calls.
We use the \textbf{SOCKET\_RW\_LEVEL} monitoring relaxation policy, described in previous work~\cite{volckaert2016secure}, to classify system calls in security-sensitive and non-sensitive ones.
\coolname{} uses four main components to achieve these goals:

\begin{enumerate}

	\item \textbf{\oldtoolname{}.} A distributed cross-process monitor (DCP-MON). \oldtoolname{}~\cite{DBLP:conf/dimva/VoulimeneasSPNL20} handles \emph{sensitive} system calls.
	\item \textbf{\disinprocessmonitor{}.} A distributed in-process monitor that is loaded into each variant's address space. \emph{Non-sensitive} system calls are forwarded to \disinprocessmonitor{} for monitoring and replication.
	\item \textbf{Connector.} A \connector{} component is assigned to every variant. Each \connector{} component communicates with \disinprocessmonitor{} through a \emph{special} shared memory segment, called the communication buffer (CB), and is responsible for transferring data to/from other \connector{} components.
	\item \textbf{\broker{}.} The syscall arbiter (SA) is a small in-kernel component that forwards \emph{sensitive} system calls to \oldtoolname{} and \emph{non-sensitive} ones to \disinprocessmonitor{}. It also restricts \disinprocessmonitor{} to execute only authenticated system calls.

\end{enumerate}

\begin{figure}[t]
	\centering
	\includegraphics[scale=0.60]{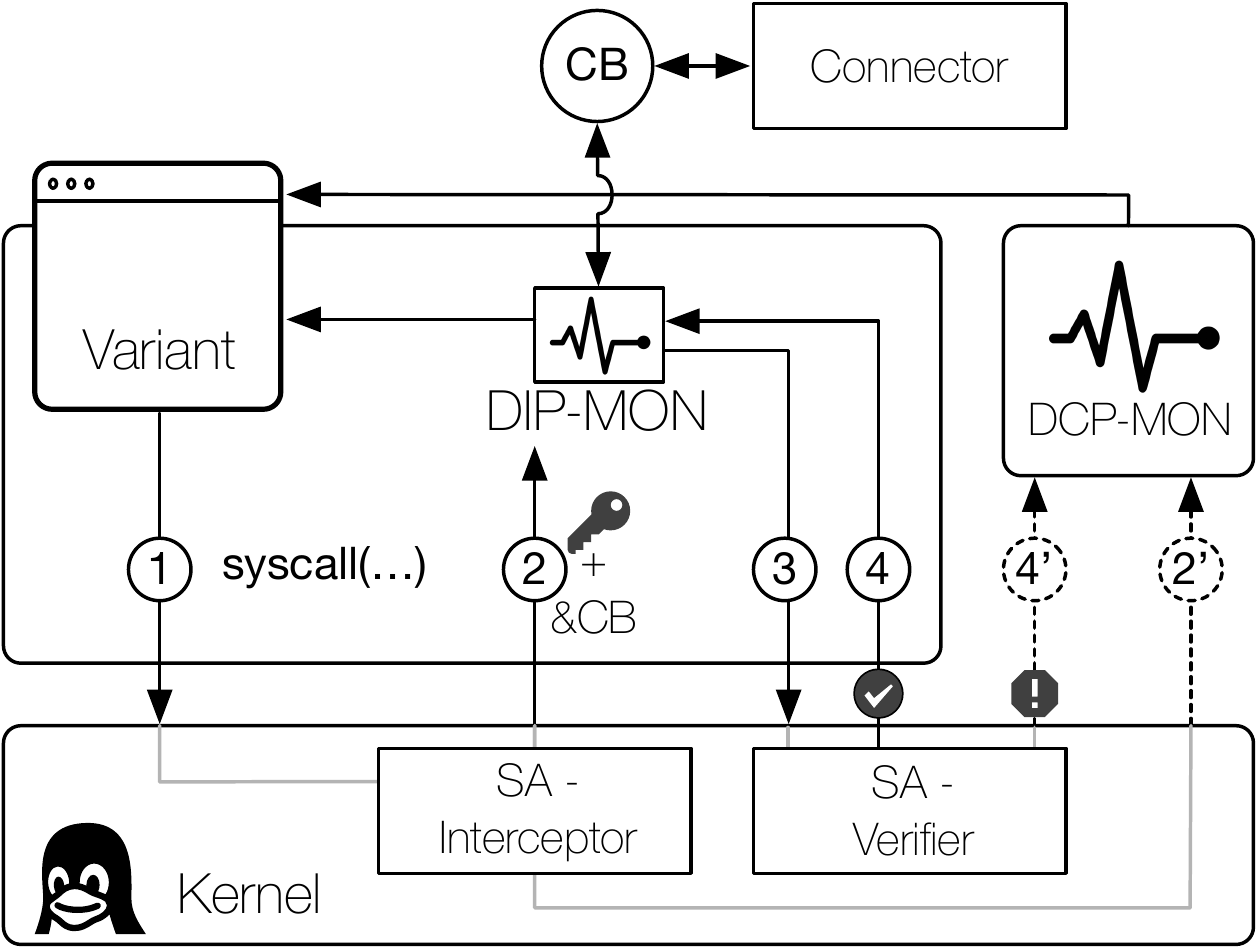}
	\caption{\coolname{'s} components and interactions.}
	\label{fig:dmon++}
\end{figure}


\noindent These four components interact with each other whenever a variant executes a system call, as shown in  \autoref{fig:dmon++}.

Our syscall arbiter intercepts system calls (\circledbw{1}) and decides if they will be handled by \disinprocessmonitor{} (\circledbw{2}) or \oldtoolname{} (\dcircledw{2\textquoteright}) depending on their security-sensitivity.
The syscall arbiter is responsible for overwriting the program counter and for forwarding non-sensitive system calls to \disinprocessmonitor{}, while \texttt{ptrace} is used to forward sensitive system calls to \oldtoolname{}.
In the former case, the syscall arbiter also creates a one-time authentication token and passes it to \disinprocessmonitor{} (\circledbw{2}) through a designated register.
Next, the \disinprocessmonitor{} performs some security checks and restarts the system call (\circledbw{3}).
The syscall arbiter permits the completion of the system call only if the authentication token is intact and the system call is issued by \disinprocessmonitor{} itself (\circledbw{4}).
In any other case, the system call is forwarded to \oldtoolname{} (\dcircledw{4\textquoteright}).\looseness=-1

We tackle the challenges described at the beginning of \autoref{sec:design} by introducing (i) a \connector{} component which acts as a proxy between \disinprocessmonitor{} and the network connection, (ii) a communication buffer stored in a shared memory segment, and (iii) a set of techniques for efficient monitoring and replication over a network connection.
The \connector{} component runs as a separate process that is not supervised by \coolname{}.
Thus, unlike \disinprocessmonitor{} itself, the \connector{} component can issue unmonitored system calls.
\disinprocessmonitor{} forwards all of its outgoing network traffic to the \connector{} component using the communication buffer, which is mapped into both processes.
It also receives incoming network traffic through this same buffer.
We also use the communication buffer to send notifications when new incoming or outgoing network traffic is available.
This setup allows us to separate the system calls we issue for monitoring and replication from the calls issued by the variants themselves.

The communication buffer is accessed by \disinprocessmonitor{} and the \connector{} component. Consequently, we need to ensure that no race conditions occur and that memory reads and writes to the communication buffer are ordered correctly. To solve these problems, we implemented and used portable synchronization primitives.


\subsection{The \disinprocessmonitor{} File Map}
\oldtoolname{} and \disinprocessmonitor{} keep their own metadata of the open file descriptors in order to ensure I/O transparency 
and optimize monitoring and replication handlers. 
We maintain a single copy of this metadata for each variant, which is mapped in each variant's address space. We refer to this metadata as the \disinprocessmonitor{} file map and we implemented it in a way similar to the communication buffer. The \disinprocessmonitor{} file map is shared between \oldtoolname{} and \disinprocessmonitor{} ensuring that both of them have a \emph{complete} view of the variant's file descriptors at any point in time. A similar mechanism was also implemented in ReMon~\cite{volckaert2016secure}. To avoid re-ordering of read and writes as well as race conditions, we implemented this using portable synchronization primitives. 

\subsection{Closing Additional Attack Surface}


Our threat model (see~\autoref{sec:threat_model}) assumes that attackers may target \disinprocessmonitor{}, the \disinprocessmonitor{} file map and the communication buffer.
We protect these components as follows.
First, we ensure that system calls that could be used to tamper with \disinprocessmonitor{} are always monitored (e.g., \texttt{mprotect}).
Second, we ensure that sensitive values like the pointer to the communication buffer never leak to memory.
To do so, we always keep sensitive values in a limited set of designated registers, and we use the \texttt{-ffixed-reg} compiler option to prevent \disinprocessmonitor{} from spilling these registers to the stack or from reusing the registers for non-sensitive values.
Third, to fully hide the location of sensitive values, while still allowing benign accesses, we ensure that the pointer to these values are only stored in kernel memory and passed to \disinprocessmonitor{} through dedicated registers \emph{only} when needed.
\disinprocessmonitor{} clobbers the sensitive registers before returning to the application's code.
Previous work implemented similar techniques and used information hiding and/or hardware protection features to protect the in-process monitor~\cite{volckaert2016secure, 10.1145/3380786.3391398}.

\subsection{Efficient Monitoring and Replication}
\label{sec:optimizations}

\coolname{} is a distributed MVX system.
Previous work has shown that most of the overhead of such a system comes from expensive network communication needed for monitoring and replication purposes and context switching caused from \texttt{ptrace}~\cite{DBLP:conf/dimva/VoulimeneasSPNL20}.
We present several techniques to reduce the performance degradation stemming from the aforementioned sources.

\subsubsection{Selective Monitoring}

ReMon~\cite{volckaert2016secure}, MvArmor~\cite{koning2016secure} and \oldtoolname{}~\cite{DBLP:conf/dimva/VoulimeneasSPNL20} categorize system calls as \emph{non-sensitive} and \emph{sensitive} ones using various security policies.
All these systems avoid monitoring the \emph{non-sensitive} system calls to minimize their run-time overhead.
In addition, ReMon~\cite{volckaert2016secure} forwards \emph{sensitive} system calls to a secure cross-process monitor and \emph{non-sensitive} system calls to a fast in-process monitor, optimizing handling of \emph{non-sensitive} system calls by avoiding context switching.
We implemented these techniques in \coolname{} and we refer to them as \emph{selective monitoring}.
Selective monitoring has two settings: strict selective monitoring in which system calls that are forwarded to \disinprocessmonitor{} are checked for equivalence and relaxed selective monitoring in which \disinprocessmonitor{} avoids monitoring of system calls that it handles.

\subsubsection{Asynchronous Replication}

The \connector{} component is implemented as a standalone process in our design. Consequently, the \disinprocessmonitor{} assigned to the leader variant could just copy the leader's system call result to the communication buffer, inform the \connector{} component and return.
This avoids stalling the leader variant; the leader variant no longer needs to wait until the network communication between the \connector{} components is finished.
Asynchronous replication is always enabled.

\subsubsection{Selective Replication}

Non-distributed MVX systems can perform replication efficiently via shared memory.
However, replication becomes expensive in a distributed setting, since network communication is needed.
We present selective replication techniques that can reduce the overhead of replication for common server applications.

\paragraph*{\textbf{Relaxed Permissive Filesystem Access}}

In previous work~\cite{DBLP:conf/dimva/VoulimeneasSPNL20}, each variant performs read-only operations on its own copy of the file, avoiding expensive replication. We find that this optimization can be extended to write operations. We extend it to all I/O operations in files that are located in the application's sub-directories. We assume that application files are in the same paths on both physical machines. Consequently, each variant could perform I/O operations on its own copy. We also assume that these files are not changed by a source other than the application itself, ensuring consistency and identical inputs. If a monitored application tries to create a file, a separate copy is created for each physical machine.

\paragraph*{\textbf{Avoiding Replication for Predictable System Call Results}}

Several system calls such as \texttt{open} and \texttt{setsockopt} are executed frequently in popular server applications. Therefore, avoiding replication for these system calls is crucial for reducing the performance overhead. \texttt{open} and similar 
system calls open and possibly create a file specified by \emph{pathname}. When successful, the system call returns an integer representing the file descriptor value that corresponds to the newly opened file. MVX systems always need to replicate the results of \texttt{open} and similar system calls, since leader and follower variants have different \emph{open} file descriptors, e.g., network sockets are only open in the leader variant; otherwise variants would get different inputs in subsequent system calls, resulting in unintended divergence. By keeping metadata of all open file descriptors,
each variant can reliably predict the next file descriptor value to be returned to the leader variant, which eliminates the need to replicate the results of \texttt{open} and similar system calls to the follower variants.\footnote{The Linux kernel returns the minimum integer value that is available for a file descriptor and it depends on the previously opened file descriptors including files, network sockets and pipes.}

Another system call that is always replicated by MVX systems is \texttt{setsockopt}. 
This system call sets options on the socket specified by the file descriptor argument, and returns 0 on success. On error, -1 is returned, and \emph{errno} is set appropriately. \coolname{} keeps metadata for all open sockets locally in each variant. Consequently, \coolname{} could reliably predict the success of \texttt{setsockopt}, by checking (using its metadata) the validity of the socket and the socket options, thereby eliminating the need to replicate the results. Note that a misprediction may lead to inconsistencies, as variants will get different system call results. Mispredictions happen only when a system call fails in one variant while succeeding in another one. We provide two configurable options in case of a system call failure: either we restart the system call until it succeeds or we terminate the variants (most applications follow one of these approaches when a system call fails).

\section{Evaluation}

We performed our experiments on two x86-64 machines running Ubuntu 16.04.5 LTS, connected using a private 100 gigabit connection between two Mellanox ConnectX Ethernet interface cards. We ran two  variants, one on each machine.

We first evaluated the performance of \coolname{} on the same microbenchmarks used in prior work~\cite{DBLP:conf/dimva/VoulimeneasSPNL20}.
For each microbenchmark, we measured the execution time under \coolname{} relative to the native execution time without \coolname{}.
\begin{figure}[t]
	\centering
	\includegraphics[scale=0.45]{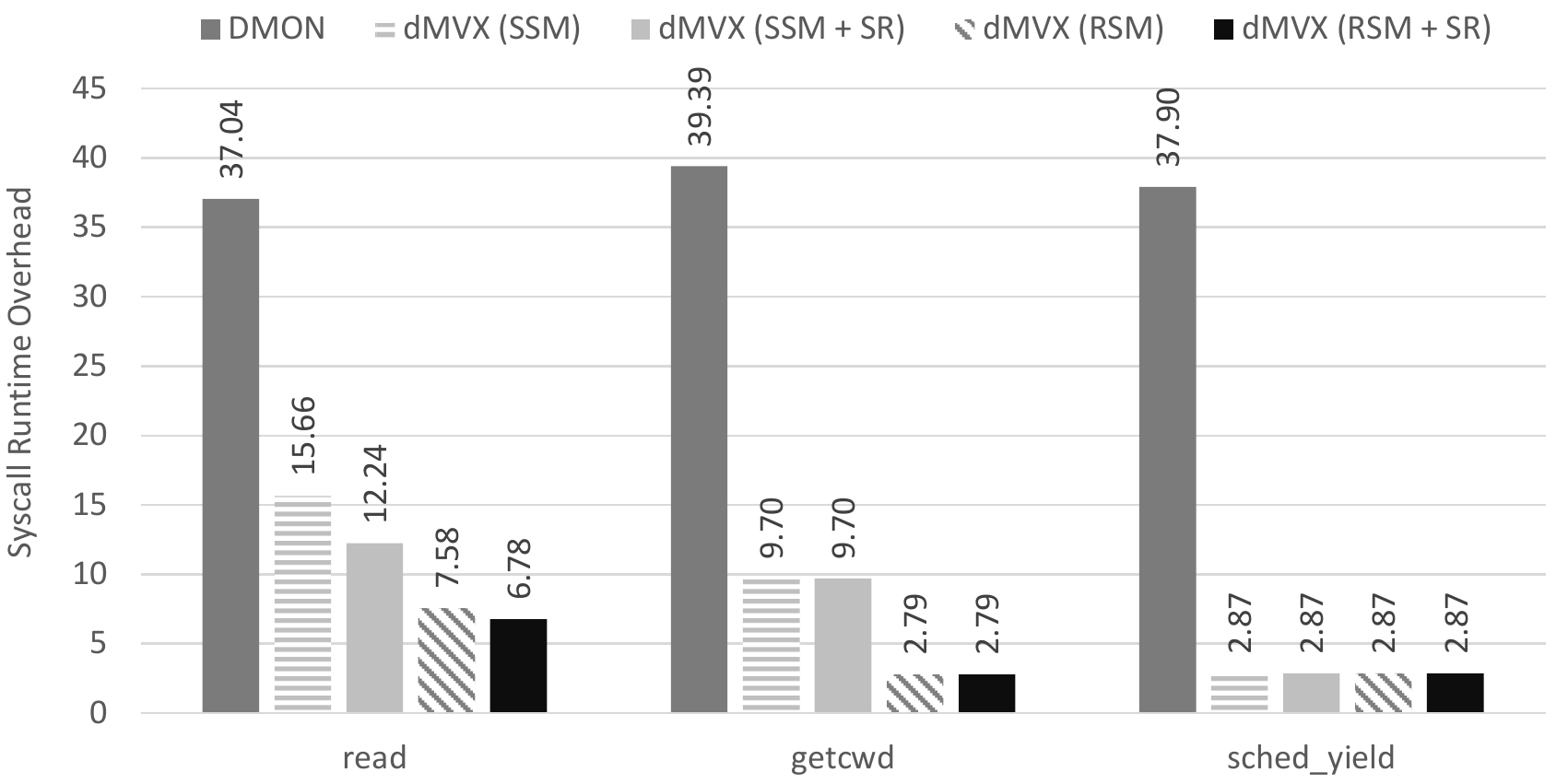}
	\caption{Microbenchmarks for 3 system calls.}
	\label{fig:microbenchmarks}
\end{figure}
\autoref{fig:microbenchmarks} shows the execution time of three representative
system calls, relative to their native execution time. We ran each benchmark
three times and took the mathematical average. Each benchmark was run with and
without our optimizations enabled. Strict and relaxed selective monitoring
appear as SSM and RSM respectively in the graphs, and selective replication
appears as SR. Our results show that the per system call overhead is
significantly reduced compared to previous
work~\cite{DBLP:conf/dimva/VoulimeneasSPNL20}, e.g., for \texttt{getcwd} the
overhead drops from 39.39$\times$ to 2.79$\times$ when all optimizations are enabled.

\begin{figure}[t]
	\centering
	\includegraphics[scale=0.45]{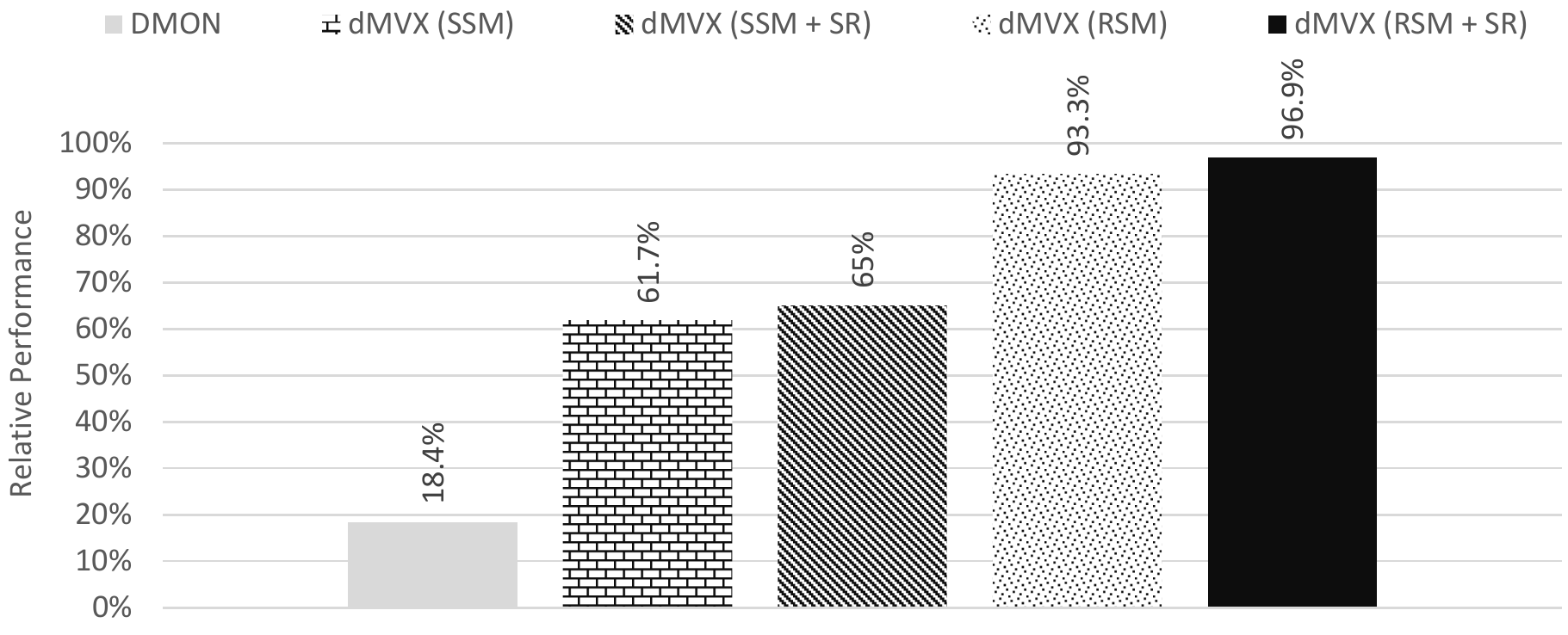}
	\caption{\texttt{lighttpd} benchmark.}
	\label{fig:serverbenchmark}
\end{figure}

We also evaluated \coolname{} on \texttt{lighttpd}, which was also used to evaluate previous work~\cite{DBLP:conf/dimva/VoulimeneasSPNL20, xiaoguang:raid20}. We used the same benchmarking tools and settings, as described in previous work~\cite{DBLP:conf/dimva/VoulimeneasSPNL20}. We measured the performance of \coolname{} relative to a vanilla \texttt{lighttpd}'s throughput. \autoref{fig:serverbenchmark} shows our results. With all of our optimizations enabled, the run-time overhead of \coolname{} on \texttt{lighttpd} drops from 443\% to just 3.1\%, a significant reduction from \oldtoolname{}~\cite{DBLP:conf/dimva/VoulimeneasSPNL20}. Furthermore, for a similar setting HeterSec introduced \char`\~50\% overhead~\cite{xiaoguang:raid20}, which is significantly higher than 3.1\%.

\section{Discussion}
\label{sec:dis}

\paragraph*{\textbf{Multi-Variant Execution}}

The design space of multi-variant execution systems has been extensively explored for security and reliability purposes~\cite{berger2006diehard,dieharder,bruschi2007diversified,cox2006n,salamat2009orchestra,volckaert2012ghumvee,koning2016secure,volckaert2016cloning,volckaert2016secure,xu2017bunshin,lu2018stopping,hosek2013safe, hosek2015varan, maurer2012tachyon, kim2015dual, kwon2016ldx, osterlund2019kmvx}. Researchers proposed techniques to address various problems that are common to MVX systems such as thread synchronization~\cite{volckaert2017taming}, consistent signal delivery~\cite{bruschi2007diversified, salamat2009orchestra}, issues with shared memory~\cite{bruschi2007diversified}, address-dependent behavior~\cite{volckaert2012ghumvee} and high overhead of security-oriented MVX systems~\cite{10.1145/3380786.3391398, volckaert2016secure}.

\paragraph*{\textbf{Diversity via ISA/ABI-Heterogeneity}}

Modern code-reuse and data-only attacks can bypass traditional software diversity techniques~\cite{larsen.etal+14} even when amplified by MVX systems due to the limited diversity that a single platform offers. Researchers proposed to leverage heterogeneous-ISA CPUs to increase the available entropy~\cite{10.1145/2872362.2872408, 10.5555/2665671.2665692}. However, these techniques require emulation, since heterogeneous-ISA platforms are not widely available yet.

Recent work in multi-variant execution places variants in heterogeneous commodity hardware, giving the illusion of a heterogeneous-ISA platform~\cite{xiaoguang:raid20, DBLP:conf/dimva/VoulimeneasSPNL20}. Researchers showed that the additional diversity, stemming from the heterogeneous-ISAs/ABIs, significantly raises the bar for code-reuse and data-only attacks.

Furthermore, ISA/ABI-heterogeneity diversifies the micro-architectural behavior of the running variants. This could potentially raise the bar for micro-architectural attacks~\cite{gruss2016rowhammer, seaborn2015exploiting, van2016drammer, 8835233, Lipp2018MeltdownRK}. The reason is that multiple building blocks of micro-architectural attacks are platform-dependent. For example, memory translation schemes and cache timing measurements are tightly coupled to the underlying architecture, making the micro-architectural attack payloads non-portable.

\section{Conclusion}

We have presented the design and implementation of 
\coolname{}, a distributed MVX system that is more efficient than prior work. \coolname{} could monitor variants that run on separate physical machines, taking advantage of the additional diversity provided by different platforms.
\coolname{} significantly reduces the overhead of prior distributed MVX systems on a real-world server benchmark by identifying \emph{expensive} MVX operations through the network that can be avoided.
Our evaluation shows that \coolname{'s} overhead on \texttt{lighttpd} is \emph{only} 3.1\%, greatly reducing the performance degradation compared to prior distributed MVX systems.


\bibliographystyle{ACM-Reference-Format}
\bibliography{sigconf}


\begin{thebibliography}{43}


\ifx \showCODEN    \undefined \def \showCODEN     #1{\unskip}     \fi
\ifx \showDOI      \undefined \def \showDOI       #1{#1}\fi
\ifx \showISBNx    \undefined \def \showISBNx     #1{\unskip}     \fi
\ifx \showISBNxiii \undefined \def \showISBNxiii  #1{\unskip}     \fi
\ifx \showISSN     \undefined \def \showISSN      #1{\unskip}     \fi
\ifx \showLCCN     \undefined \def \showLCCN      #1{\unskip}     \fi
\ifx \shownote     \undefined \def \shownote      #1{#1}          \fi
\ifx \showarticletitle \undefined \def \showarticletitle #1{#1}   \fi
\ifx \showURL      \undefined \def \showURL       {\relax}        \fi
\providecommand\bibfield[2]{#2}
\providecommand\bibinfo[2]{#2}
\providecommand\natexlab[1]{#1}
\providecommand\showeprint[2][]{arXiv:#2}

\bibitem[\protect\citeauthoryear{Abadi, Budiu, Erlingsson, and Ligatti}{Abadi
  et~al\mbox{.}}{2005}]%
        {abadi2005control}
\bibfield{author}{\bibinfo{person}{Mart{\'\i}n Abadi}, \bibinfo{person}{Mihai
  Budiu}, \bibinfo{person}{{\'Ulfar} Erlingsson}, {and} \bibinfo{person}{Jay
  Ligatti}.} \bibinfo{year}{2005}\natexlab{}.
\newblock \showarticletitle{Control-flow integrity}. In
  \bibinfo{booktitle}{\emph{Proceedings of the ACM Conference on Computer and
  Communications Security (CCS)}}.
\newblock


\bibitem[\protect\citeauthoryear{Berger and Zorn}{Berger and Zorn}{2006}]%
        {berger2006diehard}
\bibfield{author}{\bibinfo{person}{Emery~D Berger} {and}
  \bibinfo{person}{Benjamin~G Zorn}.} \bibinfo{year}{2006}\natexlab{}.
\newblock \showarticletitle{DieHard: probabilistic memory safety for unsafe
  languages}. In \bibinfo{booktitle}{\emph{Proceedings of the ACM SIGPLAN
  Conference on Programming Language Design and Implementation (PLDI)}}.
\newblock


\bibitem[\protect\citeauthoryear{Bruschi, Cavallaro, and Lanzi}{Bruschi
  et~al\mbox{.}}{2007}]%
        {bruschi2007diversified}
\bibfield{author}{\bibinfo{person}{Danilo Bruschi}, \bibinfo{person}{Lorenzo
  Cavallaro}, {and} \bibinfo{person}{Andrea Lanzi}.}
  \bibinfo{year}{2007}\natexlab{}.
\newblock \showarticletitle{Diversified process replic{\ae} for defeating
  memory error exploits}. In \bibinfo{booktitle}{\emph{IEEE Performance,
  Computing, and Communications Conference (IPCCC)}}.
\newblock


\bibitem[\protect\citeauthoryear{Cavallaro}{Cavallaro}{2007}]%
        {cavallaro_phd_thesis_2007}
\bibfield{author}{\bibinfo{person}{L Cavallaro}.}
  \bibinfo{year}{2007}\natexlab{}.
\newblock \emph{\bibinfo{title}{Comprehensive Memory Error Protection via
  Diversity and Taint-Tracking}}.
\newblock \bibinfo{thesistype}{Ph.D. Dissertation}. \bibinfo{address}{PhD
  dissertation, Universita Degli Studi Di Milano}.
\newblock


\bibitem[\protect\citeauthoryear{Checkoway, Davi, Dmitrienko, Sadeghi, Shacham,
  and Winandy}{Checkoway et~al\mbox{.}}{2010}]%
        {Shacham10}
\bibfield{author}{\bibinfo{person}{Stephen Checkoway}, \bibinfo{person}{Lucas
  Davi}, \bibinfo{person}{Alexandra Dmitrienko}, \bibinfo{person}{Ahmad{-}Reza
  Sadeghi}, \bibinfo{person}{Hovav Shacham}, {and} \bibinfo{person}{Marcel
  Winandy}.} \bibinfo{year}{2010}\natexlab{}.
\newblock \showarticletitle{Return-oriented Programming Without Returns}. In
  \bibinfo{booktitle}{\emph{Proceedings of the ACM Conference on Computer and
  Communications Security (CCS)}}.
\newblock


\bibitem[\protect\citeauthoryear{Chen, Xu, Sezer, Gauriar, and Iyer}{Chen
  et~al\mbox{.}}{2005}]%
        {dataonly}
\bibfield{author}{\bibinfo{person}{Shuo Chen}, \bibinfo{person}{Jun Xu},
  \bibinfo{person}{Emre~Can Sezer}, \bibinfo{person}{Prachi Gauriar}, {and}
  \bibinfo{person}{Ravishankar~K Iyer}.} \bibinfo{year}{2005}\natexlab{}.
\newblock \showarticletitle{Non-Control-Data Attacks Are Realistic Threats}. In
  \bibinfo{booktitle}{\emph{Proceedings of the USENIX Security Symposium}}.
\newblock


\bibitem[\protect\citeauthoryear{Cox, Evans, Filipi, Rowanhill, Hu, Davidson,
  Knight, Nguyen-Tuong, and Hiser}{Cox et~al\mbox{.}}{2006}]%
        {cox2006n}
\bibfield{author}{\bibinfo{person}{Benjamin Cox}, \bibinfo{person}{David
  Evans}, \bibinfo{person}{Adrian Filipi}, \bibinfo{person}{Jonathan
  Rowanhill}, \bibinfo{person}{Wei Hu}, \bibinfo{person}{Jack Davidson},
  \bibinfo{person}{John Knight}, \bibinfo{person}{Anh Nguyen-Tuong}, {and}
  \bibinfo{person}{Jason Hiser}.} \bibinfo{year}{2006}\natexlab{}.
\newblock \showarticletitle{N-Variant Systems: A Secretless Framework for
  Security through Diversity.}. In \bibinfo{booktitle}{\emph{USENIX Security
  Symposium}}.
\newblock


\bibitem[\protect\citeauthoryear{G{\"o}ktas, Kollenda, Koppe, Bosman,
  Portokalidis, Holz, Bos, and Giuffrida}{G{\"o}ktas et~al\mbox{.}}{2018}]%
        {bdb9f72346e94b59a6cd56b83f7c59b0}
\bibfield{author}{\bibinfo{person}{Enes G{\"o}ktas}, \bibinfo{person}{Benjamin
  Kollenda}, \bibinfo{person}{Philipp Koppe}, \bibinfo{person}{Erik Bosman},
  \bibinfo{person}{Georgios Portokalidis}, \bibinfo{person}{Thorsten Holz},
  \bibinfo{person}{Herbert Bos}, {and} \bibinfo{person}{Cristiano Giuffrida}.}
  \bibinfo{year}{2018}\natexlab{}.
\newblock \showarticletitle{Position-independent Code Reuse: On the
  Effectiveness of {ASLR} in the Absence of Information Disclosure}. In
  \bibinfo{booktitle}{\emph{IEEE European Symposium on Security and Privacy
  (EuroS\&P)}}.
\newblock


\bibitem[\protect\citeauthoryear{Gruss, Maurice, and Mangard}{Gruss
  et~al\mbox{.}}{2016}]%
        {gruss2016rowhammer}
\bibfield{author}{\bibinfo{person}{Daniel Gruss},
  \bibinfo{person}{Cl{\'e}mentine Maurice}, {and} \bibinfo{person}{Stefan
  Mangard}.} \bibinfo{year}{2016}\natexlab{}.
\newblock \showarticletitle{Rowhammer.js: A remote software-induced fault
  attack in javascript}. In \bibinfo{booktitle}{\emph{Proceedings of the
  Conference on Detection of Intrusions and Malware \& Vulnerability Assessment
  (DIMVA)}}.
\newblock


\bibitem[\protect\citeauthoryear{Hosek and Cadar}{Hosek and Cadar}{2013}]%
        {hosek2013safe}
\bibfield{author}{\bibinfo{person}{Petr Hosek} {and} \bibinfo{person}{Cristian
  Cadar}.} \bibinfo{year}{2013}\natexlab{}.
\newblock \showarticletitle{Safe software updates via multi-version execution}.
  In \bibinfo{booktitle}{\emph{Proceedings of the International Conference on
  Software Engineering (ICSE)}}.
\newblock


\bibitem[\protect\citeauthoryear{Hosek and Cadar}{Hosek and Cadar}{2015}]%
        {hosek2015varan}
\bibfield{author}{\bibinfo{person}{Petr Hosek} {and} \bibinfo{person}{Cristian
  Cadar}.} \bibinfo{year}{2015}\natexlab{}.
\newblock \showarticletitle{Varan the unbelievable: An efficient n-version
  execution framework}. In \bibinfo{booktitle}{\emph{Proceedings of the
  International Conference on Architectural Support for Programming Languages
  and Operating Systems (ASPLOS)}}.
\newblock


\bibitem[\protect\citeauthoryear{Hu, Chua, Adrian, Saxena, and Liang}{Hu
  et~al\mbox{.}}{2015}]%
        {hu2015automatic}
\bibfield{author}{\bibinfo{person}{Hong Hu}, \bibinfo{person}{Zheng~Leong
  Chua}, \bibinfo{person}{Sendroiu Adrian}, \bibinfo{person}{Prateek Saxena},
  {and} \bibinfo{person}{Zhenkai Liang}.} \bibinfo{year}{2015}\natexlab{}.
\newblock \showarticletitle{Automatic Generation of Data-Oriented Exploits.}.
  In \bibinfo{booktitle}{\emph{Proceedings of the USENIX Security Symposium}}.
\newblock


\bibitem[\protect\citeauthoryear{Hu, Shinde, Adrian, Chua, Saxena, and
  Liang}{Hu et~al\mbox{.}}{2016}]%
        {hu2016data}
\bibfield{author}{\bibinfo{person}{Hong Hu}, \bibinfo{person}{Shweta Shinde},
  \bibinfo{person}{Sendroiu Adrian}, \bibinfo{person}{Zheng~Leong Chua},
  \bibinfo{person}{Prateek Saxena}, {and} \bibinfo{person}{Zhenkai Liang}.}
  \bibinfo{year}{2016}\natexlab{}.
\newblock \showarticletitle{Data-oriented programming: On the expressiveness of
  non-control data attacks}. In \bibinfo{booktitle}{\emph{Proceedings of the
  IEEE Symposium on Security and Privacy}}.
\newblock


\bibitem[\protect\citeauthoryear{Kim, Kwon, Sumner, Zhang, and Xu}{Kim
  et~al\mbox{.}}{2015}]%
        {kim2015dual}
\bibfield{author}{\bibinfo{person}{Dohyeong Kim}, \bibinfo{person}{Yonghwi
  Kwon}, \bibinfo{person}{William~N Sumner}, \bibinfo{person}{Xiangyu Zhang},
  {and} \bibinfo{person}{Dongyan Xu}.} \bibinfo{year}{2015}\natexlab{}.
\newblock \showarticletitle{Dual execution for on the fly fine grained
  execution comparison}. In \bibinfo{booktitle}{\emph{Proceedings of the
  International Conference on Architectural Support for Programming Languages
  and Operating Systems (ASPLOS)}}.
\newblock


\bibitem[\protect\citeauthoryear{{Kocher}, {Horn}, {Fogh}, {Genkin}, {Gruss},
  {Haas}, {Hamburg}, {Lipp}, {Mangard}, {Prescher}, {Schwarz}, and
  {Yarom}}{{Kocher} et~al\mbox{.}}{2019}]%
        {8835233}
\bibfield{author}{\bibinfo{person}{P. {Kocher}}, \bibinfo{person}{J. {Horn}},
  \bibinfo{person}{A. {Fogh}}, \bibinfo{person}{D. {Genkin}},
  \bibinfo{person}{D. {Gruss}}, \bibinfo{person}{W. {Haas}},
  \bibinfo{person}{M. {Hamburg}}, \bibinfo{person}{M. {Lipp}},
  \bibinfo{person}{S. {Mangard}}, \bibinfo{person}{T. {Prescher}},
  \bibinfo{person}{M. {Schwarz}}, {and} \bibinfo{person}{Y. {Yarom}}.}
  \bibinfo{year}{2019}\natexlab{}.
\newblock \showarticletitle{Spectre Attacks: Exploiting Speculative Execution}.
  In \bibinfo{booktitle}{\emph{Proceedings of the IEEE Symposium on Security
  and Privacy}}.
\newblock


\bibitem[\protect\citeauthoryear{Koning, Bos, and Giuffrida}{Koning
  et~al\mbox{.}}{2016}]%
        {koning2016secure}
\bibfield{author}{\bibinfo{person}{Koen Koning}, \bibinfo{person}{Herbert Bos},
  {and} \bibinfo{person}{Cristiano Giuffrida}.}
  \bibinfo{year}{2016}\natexlab{}.
\newblock \showarticletitle{Secure and efficient multi-variant execution using
  hardware-assisted process virtualization}. In
  \bibinfo{booktitle}{\emph{IEEE/IFIP Conference on Dependable Systems and
  Networks (DSN)}}.
\newblock


\bibitem[\protect\citeauthoryear{Kwon, Kim, Sumner, Kim, Saltaformaggio, Zhang,
  and Xu}{Kwon et~al\mbox{.}}{2016}]%
        {kwon2016ldx}
\bibfield{author}{\bibinfo{person}{Yonghwi Kwon}, \bibinfo{person}{Dohyeong
  Kim}, \bibinfo{person}{William~Nick Sumner}, \bibinfo{person}{Kyungtae Kim},
  \bibinfo{person}{Brendan Saltaformaggio}, \bibinfo{person}{Xiangyu Zhang},
  {and} \bibinfo{person}{Dongyan Xu}.} \bibinfo{year}{2016}\natexlab{}.
\newblock \showarticletitle{{LDX}: Causality inference by lightweight dual
  execution}. In \bibinfo{booktitle}{\emph{Proceedings of the International
  Conference on Architectural Support for Programming Languages and Operating
  Systems (ASPLOS)}}.
\newblock


\bibitem[\protect\citeauthoryear{Larsen, Homescu, Brunthaler, and Franz}{Larsen
  et~al\mbox{.}}{2014}]%
        {larsen.etal+14}
\bibfield{author}{\bibinfo{person}{Per Larsen}, \bibinfo{person}{Andrei
  Homescu}, \bibinfo{person}{Stefan Brunthaler}, {and} \bibinfo{person}{Michael
  Franz}.} \bibinfo{year}{2014}\natexlab{}.
\newblock \showarticletitle{{SoK}: Automated Software Diversity}. In
  \bibinfo{booktitle}{\emph{Proceedings of the IEEE Symposium on Security and
  Privacy}}.
\newblock


\bibitem[\protect\citeauthoryear{Lipp, Schwarz, Gruss, Prescher, Haas, Fogh,
  Horn, Mangard, Kocher, Genkin, Yarom, and Hamburg}{Lipp
  et~al\mbox{.}}{2018}]%
        {Lipp2018MeltdownRK}
\bibfield{author}{\bibinfo{person}{Moritz Lipp}, \bibinfo{person}{M. Schwarz},
  \bibinfo{person}{D. Gruss}, \bibinfo{person}{Thomas Prescher},
  \bibinfo{person}{W. Haas}, \bibinfo{person}{A. Fogh}, \bibinfo{person}{Jann
  Horn}, \bibinfo{person}{S. Mangard}, \bibinfo{person}{P. Kocher},
  \bibinfo{person}{Daniel Genkin}, \bibinfo{person}{Yuval Yarom}, {and}
  \bibinfo{person}{Michael Hamburg}.} \bibinfo{year}{2018}\natexlab{}.
\newblock \showarticletitle{Meltdown: Reading Kernel Memory from User Space}.
  In \bibinfo{booktitle}{\emph{Proceedings of the USENIX Security Symposium}}.
\newblock


\bibitem[\protect\citeauthoryear{Lu, Xu, Song, Kim, and Lee}{Lu
  et~al\mbox{.}}{2018}]%
        {lu2018stopping}
\bibfield{author}{\bibinfo{person}{Kangjie Lu}, \bibinfo{person}{Meng Xu},
  \bibinfo{person}{Chengyu Song}, \bibinfo{person}{Taesoo Kim}, {and}
  \bibinfo{person}{Wenke Lee}.} \bibinfo{year}{2018}\natexlab{}.
\newblock \showarticletitle{Stopping Memory Disclosures via Diversification and
  Replicated Execution}.
\newblock \bibinfo{journal}{\emph{IEEE Transactions on Dependable and Secure
  Computing (TDSC)}} (\bibinfo{year}{2018}).
\newblock


\bibitem[\protect\citeauthoryear{Maurer and Brumley}{Maurer and
  Brumley}{2012}]%
        {maurer2012tachyon}
\bibfield{author}{\bibinfo{person}{Matthew Maurer} {and} \bibinfo{person}{David
  Brumley}.} \bibinfo{year}{2012}\natexlab{}.
\newblock \showarticletitle{{TACHYON}: Tandem execution for efficient live
  patch testing}. In \bibinfo{booktitle}{\emph{Proceedings of the USENIX
  Security Symposium}}.
\newblock


\bibitem[\protect\citeauthoryear{Nagarakatte, Zhao, Martin, and
  Zdancewic}{Nagarakatte et~al\mbox{.}}{2009}]%
        {nagarakatte2009softbound}
\bibfield{author}{\bibinfo{person}{Santosh Nagarakatte},
  \bibinfo{person}{Jianzhou Zhao}, \bibinfo{person}{Milo~M.K. Martin}, {and}
  \bibinfo{person}{Steve Zdancewic}.} \bibinfo{year}{2009}\natexlab{}.
\newblock \showarticletitle{{SoftBound}: Highly Compatible and Complete Spatial
  Memory Safety for {C}}. In \bibinfo{booktitle}{\emph{Proceedings of the ACM
  SIGPLAN Conference on Programming Language Design and Implementation
  (PLDI)}}.
\newblock


\bibitem[\protect\citeauthoryear{Nagarakatte, Zhao, Martin, and
  Zdancewic}{Nagarakatte et~al\mbox{.}}{2010}]%
        {nagarakatte2010cets}
\bibfield{author}{\bibinfo{person}{Santosh Nagarakatte},
  \bibinfo{person}{Jianzhou Zhao}, \bibinfo{person}{Milo~M.K. Martin}, {and}
  \bibinfo{person}{Steve Zdancewic}.} \bibinfo{year}{2010}\natexlab{}.
\newblock \showarticletitle{{CETS}: Compiler Enforced Temporal Safety for {C}}.
  In \bibinfo{booktitle}{\emph{International Symposium on Memory Management
  (ISMM)}}.
\newblock


\bibitem[\protect\citeauthoryear{Novark and Berger}{Novark and Berger}{2010}]%
        {dieharder}
\bibfield{author}{\bibinfo{person}{Gene Novark} {and} \bibinfo{person}{Emery~D
  Berger}.} \bibinfo{year}{2010}\natexlab{}.
\newblock \showarticletitle{{DieHarder}: Securing the Heap}. In
  \bibinfo{booktitle}{\emph{Proceedings of the ACM Conference on Computer and
  Communications Security (CCS)}}.
\newblock


\bibitem[\protect\citeauthoryear{Salamat, Jackson, Gal, and Franz}{Salamat
  et~al\mbox{.}}{2009}]%
        {salamat2009orchestra}
\bibfield{author}{\bibinfo{person}{Babak Salamat}, \bibinfo{person}{Todd
  Jackson}, \bibinfo{person}{Andreas Gal}, {and} \bibinfo{person}{Michael
  Franz}.} \bibinfo{year}{2009}\natexlab{}.
\newblock \showarticletitle{Orchestra: intrusion detection using parallel
  execution and monitoring of program variants in user-space}. In
  \bibinfo{booktitle}{\emph{Proceedings of the ACM European Conference on
  Computer Systems (EuroSys)}}.
\newblock


\bibitem[\protect\citeauthoryear{Salamat, Jackson, Gregor, and Franz}{Salamat
  et~al\mbox{.}}{2011}]%
        {salamat2011}
\bibfield{author}{\bibinfo{person}{Babak Salamat}, \bibinfo{person}{Todd
  Jackson}, \bibinfo{person}{Christian~Wimmer Gregor, Wagner}, {and}
  \bibinfo{person}{Michael Franz}.} \bibinfo{year}{2011}\natexlab{}.
\newblock \showarticletitle{Run-Time Defense against Code Injection Attacks
  using Replicated Execution}.
\newblock \bibinfo{journal}{\emph{IEEE Transactions on Dependable and Secure
  Computing (TDSC)}} (\bibinfo{year}{2011}).
\newblock


\bibitem[\protect\citeauthoryear{Seaborn and Dullien}{Seaborn and
  Dullien}{2015}]%
        {seaborn2015exploiting}
\bibfield{author}{\bibinfo{person}{Mark Seaborn} {and} \bibinfo{person}{Thomas
  Dullien}.} \bibinfo{year}{2015}\natexlab{}.
\newblock \showarticletitle{Exploiting the DRAM rowhammer bug to gain kernel
  privileges}. In \bibinfo{booktitle}{\emph{Black Hat USA}}.
\newblock


\bibitem[\protect\citeauthoryear{Shacham}{Shacham}{2007}]%
        {Shacham2007}
\bibfield{author}{\bibinfo{person}{Hovav Shacham}.}
  \bibinfo{year}{2007}\natexlab{}.
\newblock \showarticletitle{The Geometry of Innocent Flesh on the Bone:
  Return-into-libc Without Function Calls (on the x86)}. In
  \bibinfo{booktitle}{\emph{Proceedings of the ACM Conference on Computer and
  Communications Security (CCS)}}.
\newblock


\bibitem[\protect\citeauthoryear{Snow, Monrose, Davi, Dmitrienko, Liebchen, and
  Sadeghi}{Snow et~al\mbox{.}}{2013}]%
        {snow.etal+13}
\bibfield{author}{\bibinfo{person}{Kevin~Z. Snow}, \bibinfo{person}{Fabian
  Monrose}, \bibinfo{person}{Lucas Davi}, \bibinfo{person}{Alexandra
  Dmitrienko}, \bibinfo{person}{Christopher Liebchen}, {and}
  \bibinfo{person}{Ahmad{-}Reza Sadeghi}.} \bibinfo{year}{2013}\natexlab{}.
\newblock \showarticletitle{Just-In-Time Code Reuse: On the Effectiveness of
  Fine-Grained Address Space Layout Randomization}. In
  \bibinfo{booktitle}{\emph{Proceedings of the IEEE Symposium on Security and
  Privacy}}.
\newblock


\bibitem[\protect\citeauthoryear{Song, Lettner, Rajasekaran, Na, Volckaert,
  Larsen, and Franz}{Song et~al\mbox{.}}{2019}]%
        {song2019sanitizing}
\bibfield{author}{\bibinfo{person}{Dokyung Song}, \bibinfo{person}{Julian
  Lettner}, \bibinfo{person}{Prabhu Rajasekaran}, \bibinfo{person}{Yeoul Na},
  \bibinfo{person}{Stijn Volckaert}, \bibinfo{person}{Per Larsen}, {and}
  \bibinfo{person}{Michael Franz}.} \bibinfo{year}{2019}\natexlab{}.
\newblock \showarticletitle{{SoK}: Sanitizing for Security}. In
  \bibinfo{booktitle}{\emph{Proceedings of the IEEE Symposium on Security and
  Privacy}}.
\newblock


\bibitem[\protect\citeauthoryear{Van Der~Veen, Fratantonio, Lindorfer, Gruss,
  Maurice, Vigna, Bos, Razavi, and Giuffrida}{Van Der~Veen
  et~al\mbox{.}}{2016}]%
        {van2016drammer}
\bibfield{author}{\bibinfo{person}{Victor Van Der~Veen},
  \bibinfo{person}{Yanick Fratantonio}, \bibinfo{person}{Martina Lindorfer},
  \bibinfo{person}{Daniel Gruss}, \bibinfo{person}{Cl{\'e}mentine Maurice},
  \bibinfo{person}{Giovanni Vigna}, \bibinfo{person}{Herbert Bos},
  \bibinfo{person}{Kaveh Razavi}, {and} \bibinfo{person}{Cristiano Giuffrida}.}
  \bibinfo{year}{2016}\natexlab{}.
\newblock \showarticletitle{Drammer: Deterministic rowhammer attacks on mobile
  platforms}. In \bibinfo{booktitle}{\emph{Proceedings of the ACM Conference on
  Computer and Communications Security (CCS)}}.
\newblock


\bibitem[\protect\citeauthoryear{Venkat, Shamasunder, Shacham, and
  Tullsen}{Venkat et~al\mbox{.}}{2016}]%
        {10.1145/2872362.2872408}
\bibfield{author}{\bibinfo{person}{Ashish Venkat}, \bibinfo{person}{Sriskanda
  Shamasunder}, \bibinfo{person}{Hovav Shacham}, {and} \bibinfo{person}{Dean~M
  Tullsen}.} \bibinfo{year}{2016}\natexlab{}.
\newblock \showarticletitle{Hipstr: Heterogeneous-ISA program state
  relocation}. In \bibinfo{booktitle}{\emph{Proceedings of the International
  Conference on Architectural Support for Programming Languages and Operating
  Systems (ASPLOS)}}.
\newblock


\bibitem[\protect\citeauthoryear{Venkat and Tullsen}{Venkat and
  Tullsen}{2014}]%
        {10.5555/2665671.2665692}
\bibfield{author}{\bibinfo{person}{Ashish Venkat} {and}
  \bibinfo{person}{Dean~M. Tullsen}.} \bibinfo{year}{2014}\natexlab{}.
\newblock \showarticletitle{Harnessing ISA Diversity: Design of a
  Heterogeneous-ISA Chip Multiprocessor}. In
  \bibinfo{booktitle}{\emph{International Symposium on Computer Architecture
  (ISCA)}}.
\newblock


\bibitem[\protect\citeauthoryear{Volckaert, Coppens, and De~Sutter}{Volckaert
  et~al\mbox{.}}{2016a}]%
        {volckaert2016cloning}
\bibfield{author}{\bibinfo{person}{Stijn Volckaert}, \bibinfo{person}{Bart
  Coppens}, {and} \bibinfo{person}{Bjorn De~Sutter}.}
  \bibinfo{year}{2016}\natexlab{a}.
\newblock \showarticletitle{Cloning your gadgets: Complete ROP attack immunity
  with multi-variant execution}.
\newblock \bibinfo{journal}{\emph{IEEE Transactions on Dependable and Secure
  Computing (TDSC)}} (\bibinfo{year}{2016}).
\newblock


\bibitem[\protect\citeauthoryear{Volckaert, Coppens, De~Sutter, De~Bosschere,
  Larsen, and Franz}{Volckaert et~al\mbox{.}}{2017}]%
        {volckaert2017taming}
\bibfield{author}{\bibinfo{person}{Stijn Volckaert}, \bibinfo{person}{Bart
  Coppens}, \bibinfo{person}{Bjorn De~Sutter}, \bibinfo{person}{Koen
  De~Bosschere}, \bibinfo{person}{Per Larsen}, {and} \bibinfo{person}{Michael
  Franz}.} \bibinfo{year}{2017}\natexlab{}.
\newblock \showarticletitle{Taming parallelism in a multi-variant execution
  environment}. In \bibinfo{booktitle}{\emph{Proceedings of the ACM European
  Conference on Computer Systems (EuroSys)}}.
\newblock


\bibitem[\protect\citeauthoryear{Volckaert, Coppens, Voulimeneas, Homescu,
  Larsen, De~Sutter, and Franz}{Volckaert et~al\mbox{.}}{2016b}]%
        {volckaert2016secure}
\bibfield{author}{\bibinfo{person}{Stijn Volckaert}, \bibinfo{person}{Bart
  Coppens}, \bibinfo{person}{Alexios Voulimeneas}, \bibinfo{person}{Andrei
  Homescu}, \bibinfo{person}{Per Larsen}, \bibinfo{person}{Bjorn De~Sutter},
  {and} \bibinfo{person}{Michael Franz}.} \bibinfo{year}{2016}\natexlab{b}.
\newblock \showarticletitle{Secure and Efficient Application Monitoring and
  Replication.}. In \bibinfo{booktitle}{\emph{Proceedings of the USENIX Annual
  Technical Conference (ATC)}}.
\newblock


\bibitem[\protect\citeauthoryear{Volckaert, De~Sutter, De~Baets, and
  De~Bosschere}{Volckaert et~al\mbox{.}}{2012}]%
        {volckaert2012ghumvee}
\bibfield{author}{\bibinfo{person}{Stijn Volckaert}, \bibinfo{person}{Bjorn
  De~Sutter}, \bibinfo{person}{Tim De~Baets}, {and} \bibinfo{person}{Koen
  De~Bosschere}.} \bibinfo{year}{2012}\natexlab{}.
\newblock \showarticletitle{{GHUMVEE}: efficient, effective, and flexible
  replication}. In \bibinfo{booktitle}{\emph{International Symposium on
  Foundations and Practice of Security (FPS)}}.
\newblock


\bibitem[\protect\citeauthoryear{Voulimeneas, Song, Parzefall, Na, Larsen,
  Franz, and Volckaert}{Voulimeneas et~al\mbox{.}}{2020}]%
        {DBLP:conf/dimva/VoulimeneasSPNL20}
\bibfield{author}{\bibinfo{person}{Alexios Voulimeneas},
  \bibinfo{person}{Dokyung Song}, \bibinfo{person}{Fabian Parzefall},
  \bibinfo{person}{Yeoul Na}, \bibinfo{person}{Per Larsen},
  \bibinfo{person}{Michael Franz}, {and} \bibinfo{person}{Stijn Volckaert}.}
  \bibinfo{year}{2020}\natexlab{}.
\newblock \showarticletitle{Distributed Heterogeneous N-Variant Execution}. In
  \bibinfo{booktitle}{\emph{Proceedings of the Conference on Detection of
  Intrusions and Malware \& Vulnerability Assessment (DIMVA)}}.
\newblock


\bibitem[\protect\citeauthoryear{Wahbe, Lucco, Anderson, and Graham}{Wahbe
  et~al\mbox{.}}{1993}]%
        {wahbe1994efficient}
\bibfield{author}{\bibinfo{person}{Robert Wahbe}, \bibinfo{person}{Steven
  Lucco}, \bibinfo{person}{Thomas~E Anderson}, {and} \bibinfo{person}{Susan~L
  Graham}.} \bibinfo{year}{1993}\natexlab{}.
\newblock \showarticletitle{Efficient software-based fault isolation}. In
  \bibinfo{booktitle}{\emph{ACM Symposium on Operating Systems Principles
  (SOSP)}}.
\newblock


\bibitem[\protect\citeauthoryear{Wang, Yeoh, Lyerly, Olivier, Kim, and
  Ravindran}{Wang et~al\mbox{.}}{2020a}]%
        {xiaoguang:raid20}
\bibfield{author}{\bibinfo{person}{Xiaoguang Wang}, \bibinfo{person}{SengMing
  Yeoh}, \bibinfo{person}{Robert Lyerly}, \bibinfo{person}{Pierre Olivier},
  \bibinfo{person}{Sang-Hoon Kim}, {and} \bibinfo{person}{Binoy Ravindran}.}
  \bibinfo{year}{2020}\natexlab{a}.
\newblock \showarticletitle{A Framework for Software Diversification with {ISA}
  Heterogeneity}. In \bibinfo{booktitle}{\emph{"International Symposium on
  Research in Attacks, Intrusions and Defenses (RAID)"}}.
\newblock


\bibitem[\protect\citeauthoryear{Wang, Yeoh, Olivier, and Ravindran}{Wang
  et~al\mbox{.}}{2020b}]%
        {10.1145/3380786.3391398}
\bibfield{author}{\bibinfo{person}{Xiaoguang Wang}, \bibinfo{person}{SengMing
  Yeoh}, \bibinfo{person}{Pierre Olivier}, {and} \bibinfo{person}{Binoy
  Ravindran}.} \bibinfo{year}{2020}\natexlab{b}.
\newblock \showarticletitle{Secure and Efficient In-Process Monitor (and
  Library) Protection with Intel MPK}. In \bibinfo{booktitle}{\emph{European
  Workshop on System Security (EuroSec)}}.
\newblock


\bibitem[\protect\citeauthoryear{Xu, Lu, Kim, and Lee}{Xu
  et~al\mbox{.}}{2017}]%
        {xu2017bunshin}
\bibfield{author}{\bibinfo{person}{Meng Xu}, \bibinfo{person}{Kangjie Lu},
  \bibinfo{person}{Taesoo Kim}, {and} \bibinfo{person}{Wenke Lee}.}
  \bibinfo{year}{2017}\natexlab{}.
\newblock \showarticletitle{Bunshin: compositing security mechanisms through
  diversification}. In \bibinfo{booktitle}{\emph{Proceedings of the USENIX
  Annual Technical Conference (ATC)}}.
\newblock


\bibitem[\protect\citeauthoryear{Österlund, Koning, Olivier, Barbalace, Bos,
  and Giuffrida}{Österlund et~al\mbox{.}}{2019}]%
        {osterlund2019kmvx}
\bibfield{author}{\bibinfo{person}{Sebastian Österlund}, \bibinfo{person}{Koen
  Koning}, \bibinfo{person}{Pierre Olivier}, \bibinfo{person}{Antonio
  Barbalace}, \bibinfo{person}{Herbert Bos}, {and} \bibinfo{person}{Cristiano
  Giuffrida}.} \bibinfo{year}{2019}\natexlab{}.
\newblock \showarticletitle{{kMVX}: Detecting Kernel Information Leaks with
  Multi-variant Execution}. In \bibinfo{booktitle}{\emph{Proceedings of the
  International Conference on Architectural Support for Programming Languages
  and Operating Systems (ASPLOS)}}.
\newblock


\end{thebibliography}

\end{document}